# Fluid Tunnel Research for Challenges of Urban Climate

Yongling Zhao [a], Lup Wai Chew [b], Yifan Fan [c, d], Christof Gromke [e], Jian Hang [f], Yichen Yu [g], Alessio Ricci [h, i], Yan Zhang [c, d], Yunpeng Xue [j], Sofia Fellini [k], Parham A. Mirzaei [l], Naiping Gao [m], Matteo Carpentieri [n], Pietro Salizzoni [k], Jianlei Niu [g], Jan Carmeliet [a]

[a] *Department of Mechanical and Process Engineering, ETH Zürich, Switzerland*
[b] *Department of the Built Environment, National University of Singapore, Singapore*
[c] *College of Civil Engineering and Architecture, Zhejiang University, Zhejiang University, China*
[d] *International Research Center for Green Building and Low-Carbon City, Zhejiang University, China*
[e] *Institute for Hydromechanics, Karlsruhe Institute of Technology, Germany*
[f] *School of Atmospheric Sciences, Sun Yat-Sen University, China*
[g] *Department of Building Environment and Energy Engineering, Hong Kong Polytechnic University, Hong Kong, China*
[h] *University School for Advanced Studies IUSS, Italy*
[i] *Eindhoven University of Technology, Netherlands*
[j] *Future Resilient Systems, Singapore-ETH Centre, ETH Zurich, Singapore*
[k] *Univ Lyon, INSA Lyon, CNRS, Ecole Centrale de Lyon, Univ Claude Bernard Lyon 1, France*
[l] *Architecture & Built Environment Department, University of Nottingham, United Kingdom*
[m] *Department of Mechanical Engineering, Tongji University, China*
[n] *Department of Mechanical Engineering Sciences, University of Surrey, United Kingdom*

E-mail addresses: yozhao@ethz.ch (Y. Zhao), lupwai@nus.edu.sg (LW. Chew), yifanfan@zju.edu.cn (Y. Fan), christof-bernhard.gromke@kit.edu (C. Gromke), hangj3@mail.sysu.edu.cn (J. Hang), yichen.yu@polyu.edu.hk (Y.Yu), a.ricci@tue.nl (A. Ricci), 12012029@zju.edu.cn (Y. Zhang), yunpeng.xue@sec.ethz.ch (Y.Xue), sofia.fellini@ec-lyon.fr (S.Fellini), Parham.Mirzaei_Ahranjani@nottingham.ac.uk (PA. Mirzaei), gaonaiping@tongji.edu.cn (NP. Gao), m.carpentieri@surrey.ac.uk (M. Carpentieri), pietro.salizzoni@ec-lyon.fr (P. Salizzoni), jian-lei.niu@polyu.edu.hk (J.Niu), cajan@ethz.ch (J.Carmeliet)

### *Nomenclature*

| | | |
|---|---|---|
| $Re$ | Reynolds number | - |
| $Ri$ | Richardson number | - |
| $Ro$ | Rossby number | - |
| $Sc$ | Schmidt number | - |
| $Gr$ | Grashof number | - |
| $Fr$ | Froude number | - |
| $Pe$ | Peclet number | - |
| $M$ | geometric scale factor | - |
| $U$ | freestream fluid velocity | m s$^{-1}$ |
| $U_f$ | bulk flow speed | m s$^{-1}$ |
| $F_d$ | extracted momentum (drag) | kg m s$^{-1}$ |
| $F_f$ | momentum of the approach flow | kg m s$^{-1}$ |
| $p_{dyn}$ | dynamic pressure | Pa |
| $p_{lw}$ | static pressure leeward | Pa |
| $p_{ww}$ | static pressure windward | Pa |
| $\Delta p_{st}$ | difference in static pressure windward and leeward | Pa |
| $N$ | buoyancy frequency | s$^{-1}$ |
| $T$ | temperature | K |
| $\Delta T$ | temperature difference between the heat source and the ambient | K |
| $H$ | length scale | m |
| $z$ | vertical coordinate | m |
| $d$ | porous sample thickness in streamwise direction | m |
| $g$ | gravitational acceleration | m s$^{-2}$ |
| $g'$ | reduced gravity | m s$^{-2}$ |

### *Greek letters*

| | | |
|---|---|---|
| $\beta$ | fluid thermal expansion rate | K$^{-1}$ |
| $\theta$ | potential temperature | K |
| $\lambda$ | pressure loss coefficient | m$^{-1}$ |
| $\nu$ | kinematic viscosity | m$^2$ s$^{-1}$ |
| $\rho$ | fluid density | kg m$^{-3}$ |

**Abstract**
Experimental investigations using wind and water tunnels have long been a staple of fluid mechanics research for a large number of applications. These experiments often single out a specific physical process to be investigated, while studies involving multiscale and multi-physics processes are rare due to the difficulty and complexity in the experimental setup. In the era of climate change, there is an increasing interest in innovative experimental studies in which fluid (wind and water) tunnels are employed for modelling multiscale, multi-physics phenomena of the urban climate. High-quality fluid tunnel measurements of urban-physics related phenomena are also much needed to facilitate the development and validation of advanced multi-physics numerical models. As a repository of knowledge in modelling these urban processes, we cover fundamentals, recommendations and guidelines for experimental design, recent advances and outlook on eight selected research areas, including (i) thermal buoyancy effects of urban airflows, (ii) aerodynamic and thermal effects of vegetation, (iii) radiative and convective heat fluxes over urban materials, (iv) influence of thermal stratification on land-atmosphere interactions, (v) pollutant dispersion, (vi) indoor and outdoor natural ventilation, (vii) wind thermal comfort, and (viii) urban winds over complex urban sites. Further, three main challenges, i.e., modelling of multi-physics, modelling of anthropogenic processes, and combined use of fluid tunnels, scaled outdoor and field measurements for urban climate studies, are discussed.

## 1. Introduction

Wind tunnels have been an indispensable tool in advancing mankind's technology, from Wright Brothers' Flyer to Neil Armstrong's statement "One small step for man, one giant leap for mankind" during Apollo 11 moon mission. Even with today's advanced computational power and the popularity of computational fluid dynamics (CFD), wind tunnel experiments are still widely used due to their good capability and feasibility and are still considered to be not replaceable in fluid mechanics research, especially dealing with complex flow mechanisms. While the contribution of wind tunnels in rocket science (literally) is well known, this experimental approach is essential in other areas of research, including wind engineering, urban physics, sports engineering, and many others. Water tunnels serve the same purpose in experimental fluid mechanics and offer some advantages for studies of urban climate, for instance, the possibility in measuring non-isothermal flow field at high spatial resolution and reducing model sizes because of higher viscosity of water compared to air. Both air and water are fluids, and the physics is governed by the same equations, namely the Navier-Stokes equations. Therefore, we use the term "fluid tunnel" to cover both wind and water tunnels in this paper.

Fluid tunnel experiments for urban climate research adopt scaled-down models due to the limitation of the dimensions of the fluid tunnel test section. One commonly asked question is the size of a fluid tunnel required for the modelling of a realistic, full-scale urban climate problem (Meroney, 2016). Are we able to capture the same physics at full scale using building models with a scale-down ratio of 1:10? What about scales of 1:100, 1:1000 or beyond? Similarities or dimensional analysis in fluid mechanics can provide the key dimensionless parameters important for a specific application, but often these dimensionless parameters cannot be matched in fluid tunnels. Therefore, care should be taken while applying fluid tunnel experiments results to full-scale applications. CFD can complement this weakness of scaling mismatch in fluid tunnel experiments by means of full-scale simulations (Blocken, 2014). This raises another commonly asked question: Why do we still use fluid tunnels despite the great advance in CFD and computational resources (Meroney, 2016)? The answer could be that the mutual and complementary use of these techniques guarantees an enhanced performance of both by leading to a better understanding and interpretation of the underlying physics under study(Murakami, 1990, Stathopoulos, 1997, Li et al., 2006, Blocken, 2014). It is worth noting that parameterisations of thermal and vapour fluxes at boundaries of complex geometries remain a challenge for CFD. There are many high-quality reviews and guidelines for the use of CFD in urban physics and more specifically in urban climate (e.g. Franke et al., 2007, Tominaga et al., 2008, Britter and Schatzmann, 2010, Blocken and Gualtieri, 2012, García-Sánchez et al., 2018), but such reviews and guidelines are lacking for the applications of fluid tunnels in urban climate analysis. Therefore, we believe this paper is timely to address the capability and challenges of fluid tunnel applications in urban climate.

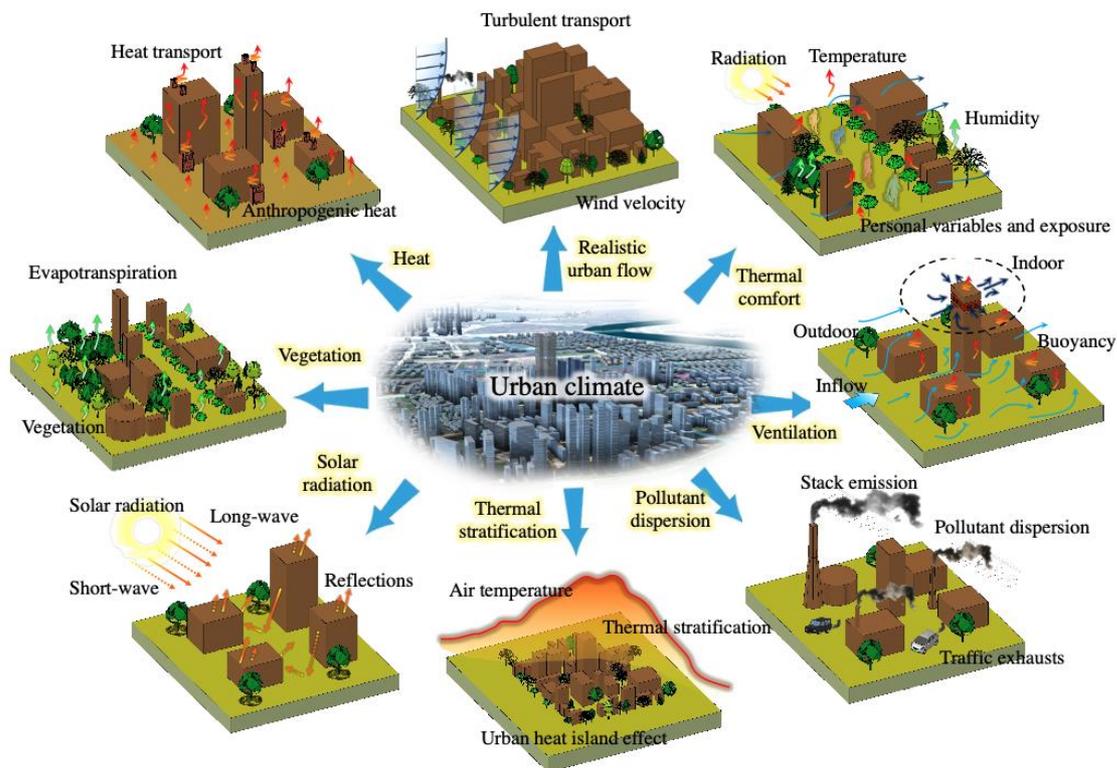

**Fig. 1**. Fluid tunnel modelling capabilities for physical processes of the urban climate.

The working principle of fluid tunnels is rather simple, where fluid is driven in bulk by fans or pumps to generate a (usually steady or quasi-steady) flow across the test section. These two ways of generating flows have the advantage of easy and accurate control of the flow rate (and hence velocity) and the use of wire mesh, screens, and honeycombs can reduce undesirable turbulence of the generated flow (e.g. Groth and Johansson, 1988, Kulkarni et al., 2011). In addition to a basic setup of a fluid tunnel experiment, three major requirements for urban climate modelling in fluid tunnels are: (i) generation and development of a desired atmospheric boundary layer (ABL), (ii) generation of heated fluxes on/from objects of interest, and (iii) generation of scalar transport. First, to achieve a fully developed approach neutral ABL flow following a log-law or power law (Stull, 1988), a roughness fetch and vortex generators can be arranged upstream of the test section (e.g. Castro and Robins, 1977, Chew et al., 2017, Zhang et al., 2017, Catarelli et al., 2020). Second, to simulate heated surfaces, for example, streets and exterior building walls heated up by solar radiation, heating elements need to be integrated into the fluid tunnel without obstructing the flows (e.g. Allegrini et al., 2013, Cui et al., 2016,

Zhao et al., 2022b). Third, modelling of scalar transport, for example pollutant dispersion, requires the release (and purging) of tracer gas or particles in the fluid tunnels (e.g. Meroney et al., 1996, Liu et al., 2010, Gromke and Ruck, 2012, Fellini et al., 2022).

When these requirements are fulfilled, fluid tunnels have the capability to realistically reproduce and model multi-physic processes of the urban climate. As depicted in Fig. 1, the eight key multi-physics research areas commonly studied in fluid tunnels are covered in this paper, though some other applications of fluid tunnels, e.g., modelling of extreme events (such as tornadoes and downburst winds), are not covered. The eight research topics are summarized below and the requirements for their fluid tunnel modelling are reported in Table A1 (Appendix A).

- Modelling of thermal buoyancy effects (Section 2.1)
- Modelling of vegetation (Section 2.2)
- Modelling of solar radiation (Section 2.3)
- Modelling of thermal stratification (Section 2.4)
- Modelling of pollutant dispersion (Section 2.5)
- Modelling of indoor and outdoor natural ventilation (Section 2.6)
- Modelling of outdoor wind thermal comfort (Section 2.7)
- Modelling of urban flow over complex urban sites (Section 2.8).

Solar radiation is the main source controlling the urban surface energy budget and driving many processes in built environments. The presence of thermal buoyancy effects in built environments is in fact largely due to solar radiation adsorption/release and anthropogenic heat generation. Thermal stratification may further establish when the buoyancy effects develop differently along the height direction, which affects wind and heat transfer in cities and also thermal comfort of residents. To mitigate excessive urban heat, urban vegetation as a means to provide shade, transpirative cooling, and modification of mean flow (wind) has become popular in many cities. The presence of vegetation and buoyancy effects in turn leads to more complex pollutant dispersion in cities. As a result, indoor and outdoor ventilation for venues in realistic urban sites have to be understood from thermal comfort and air quality point of view.

The use of fluid tunnel experiments to model the urban climate in realistic urban sites links the research to applications and implementation in the real world, including urban planning and policy making. Each of these selected research areas will be discussed in Section 2.1 – 2.8. In each section, the fundamental considerations will be firstly provided, followed by recommendations for the design of experiments, and recent advances and outlook. The remainder of the paper is organized as follows: Section 2 describes the main advances in fluid tunnel modelling of multi-physics of the urban climate; Section 3 focuses on three main challenges for fluid tunnel modelling of urban climate; Section 4 closes the paper with conclusions and remarks.

## 2. Advances in modelling of multi-physics of urban climate
### 2.1 Modelling of thermal buoyancy effects
*2.1.1 Fundamental considerations*

Urban climate is dominated by many physical processes involving thermal buoyancy. For example, urban wind flowing over asphalt pavement, building facades or roofs at high surface temperature due to solar radiation adsorption is heated up in summer daytime and thus gains buoyancy through convective heat transfer. The thermal buoyancy of urban airflow may play a vital role in pollutant dispersion, heat removal, thermal comfort of residents, etc (e.g. Dallman et al., 2014, Mei and Yuan, 2022, Mouzourides et al., 2022).

To characterise the urban airflow involving convective heat transfer from urban surfaces (e.g., ground, facades, etc.), the overall process can be regarded as an approximation of a Poiseuille-Rayleigh-Bénard-type flow, where the approaching wind can be characterised as a Poiseuille flow and the heat release from urban surfaces (grounds, buildings, roofs, etc) can be approximated as a Rayleigh-Bénard-type flow. The ratio between buoyancy and shear forcing of the incoming wind can be quantified by the bulk Richardson number ($Ri$), which can be defined in Eq.(2.1-1) (e.g. Chew et al., 2018b, Zhao et al., 2022b):

$$Ri = \frac{Gr}{Re^2} = \frac{g\beta H^3 \Delta T/\nu^2}{(U^2 H/\nu)^2} = \frac{g\beta H \Delta T}{U^2} \qquad (2.1\text{-}1)$$

where $Gr$ is the Grashof number characterising the buoyancy effect, $Re$ is the Reynolds number reflecting the shear effect. $H$ is the length scale of heat source that releases heat to ambient fluid (e.g. air), $\Delta T$ is the temperature difference between the heat source and the ambient, $\beta$ is the thermal expansion coefficient of fluid (e.g. air), $g$ is the acceleration due to gravity, $\nu$ is the fluid kinematic viscosity, and $U$ is the freestream fluid velocity. When the hydrostatic pressure variation of urban airflow is considerable, potential temperature is commonly used in the calculation of $Ri$, instead of using the absolute temperature.

*2.1.2 Recommendations for the design of fluid tunnel experiments*

Fluid tunnels and heated building models have been used extensively to generate and study buoyancy effects in urban airflows at reduced scales (e.g. Allegrini et al., 2014, Tsalicoglou et al., 2020). An example experimental setup is shown in Fig. 2a. In design of experiments, particular attention needs to be paid to the control of heating of the models, in additional to considering well-established requirements for the blockage ratio of measurement section (Jeong et al., 2018) and proper generation of the boundary layer flow profile (Catarelli et al., 2020).

Heating of model surfaces can be designed in two ways, that is, constant surface temperature (Zhao et al., 2021) or constant heat flux (Gaheen et al., 2021). The implementation of constant surface temperature can be achieved using electronic heating pads (Mouzourides et al., 2022) or by circulating heated water inside the models from a water bath (Shah et al., 2018). The implementation of constant heat flux can be achieved by using a heating power control. As the convective heat transfer coefficient of a building model surface could vary significantly under different wind conditions, the heating capacity of the electronic heater or water bath has to be chosen according to the desired surface temperature and the maximum convective heat transfer coefficient. Uniformity of surface temperatures needs to be ensured as much as possible prior to measurements, though small spatial variations could still exist due to limited control precision of heating for a large heat transfer surface.

For temperature measurements, multiple thermocouples can be placed in a rack on the velocity measurement plane to allow quasi-temperature field measurements at low spatial resolution. While this approach is straightforward to implement, it does not allow non-intrusive and simultaneous velocity and temperature field measurements, and therefore the temperature measurement needs to be performed separately.

The design of experiments also needs to facilitate the measurements. For velocity field measurements, motorized multi-dimensional stages may be used to allow efficient multiple field-of-view (FOV) measurements where lasers and the camera have to be moved in a synchronized way (Li et al., 2021). Depending on the laser intensity and optical access to the measurement plane, a mirror might be needed to enhance local laser intensity (Mouzourides et al., 2022).

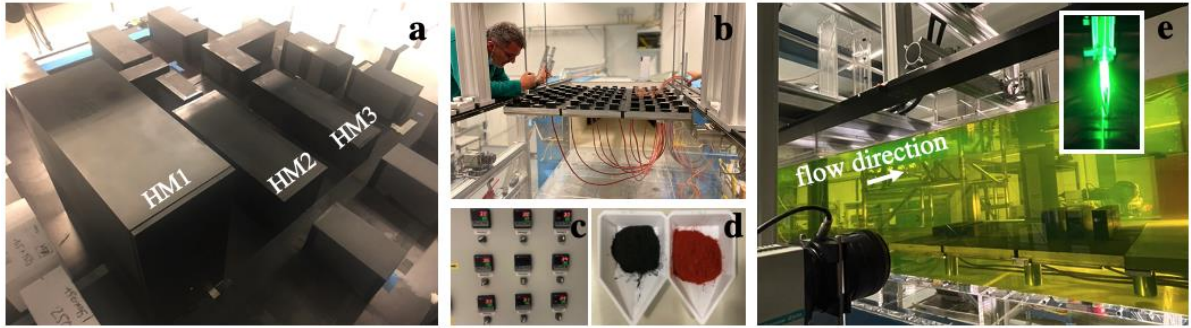

**Fig. 2**. Wind and water tunnel measurements of buoyancy effects of urban flows. (a) Wind tunnel setup showing electrical heated models (HM1-3); (b-e) water tunnel setup showing (b) conductive models on heating plates, (c) individual controllers of heating plates, (d) fluorescence chlorophyll / Uranine, and (e) optical setup.

*2.1.3 Recent advances and outlook*
Despite the complexity mentioned above, fluid tunnel experiments remain a good option for studying non-isothermal flows or thermal buoyancy effects in urban climate processes, given their benefits in providing an approaching flow of desired profile, flexibility in setting up models, established optical setup and measurements, etc. Off-shelf equipment for quasi-field temperature measurements needs to be engineered and offered, as an auxiliary to well-established particle image velocimetry (PIV) equipment.

Water tunnel measurements, as a promising approach, provide the opportunity to perform simultaneous velocity and temperature field measurements using PIV and laser-induced fluorescence (LIF)(Zhao et al., 2022b). An example setup is shown in Fig. 2b-e. For LIF, the fluorescence intensity measured at every pixel of the image is expected to have a linear relationship between the local laser intensity and temperature. The main challenges are the non-uniform spatial distribution of the laser sheet intensity, fluctuations of the laser intensity, and errors due to dye concentration and dye absorption (Vanderwel and Tavoularis, 2014, Zhao et al., 2022b). Uncertainties of the thermal couples used to calibrate the LIF results also limit the accuracy of the determined temperature field.

Conductive models in water tunnels can be placed onto heating power-controlled heating plates (Fig. 2b-c) to mimic heat sources in urban climate. Non-toxic fluorescence, such as chlorophyll and Uranine (Fig. 2d), can be used for LIF measurements. LIF measurements using 1-color/1-dye or 2-color/2-dye can be adopted, depending on the desired temperature resolution(Yen et al., 2016, Zhao et al., 2022b).

## 2.2 Modelling of vegetation
*2.2.1 Fundamental considerations*
Flow inside and past vegetation is complex. Vegetation elements (leaves or needles, twigs, branches, trunks) generate local boundary layers and wakes which interact with each other and thereby forming shear layers and other intricate flow structures. The fundamental physical phenomena occurring in the flow through vegetation are (i) extraction of momentum due to aerodynamic resistance, (ii) conversion of mean into turbulence kinetic energy, and (iii) break-up of larger-scale turbulent motions into smaller-scale ones and in this way short-circuiting the eddy cascade (Shaw, 1985).

In reduced-scale fluid tunnel studies of the built and natural environment, typically the aerodynamic load on and the modification of the flow and dispersion field in the surrounding of vegetation are of interest (Fig. 3). Hence, scaling and similarity considerations should be directed on aerodynamic resistance (drag) and permeability. While scaling of the aerodynamic resistance ensures similarity in the extraction of momentum from the flow, it does not ensure similarity for the kinematics of the flow. The latter requires scaling of the permeability which implies the partitioning of flow going through and around the vegetation. This is pivotal for the major vegetation-induced turbulent flow structures including the characteristics of the recirculation in the lee. The break-up of turbulent motions and the short-circuiting of the eddy cascade are inherently complied with, however, only in a qualitative manner. A rigorous representation of the short-circuiting of the eddy cascade and all involved scales of turbulent motions eludes scaling. This is since typical $Re$ numbers in reduced-scale experiments are two orders of magnitude smaller compared to the real scale and due to complex non-linear interactions in the vortex decay mechanism.

*2.2.2 Recommendations for the design of fluid tunnel experiments*

For the modelling of vegetation in reduced-scale wind tunnel studies, it is recommended to utilize prefabricated plastic-based open porous foams. Such foams are commercially available from several manufacturers with various porosities denoted by PPI-x, where PPI stands for 'pores per inch' and x is their count. The foams are available for with $7 < x < 100$ and can be processed e.g. with a knife or scissors to reproduce the contour of the vegetation under consideration (Fig. 3). Alternatively, open porous objects can be self-made as clusters of interwoven filament- or stripe-like components as successfully applied in previous works (Gromke and Ruck, 2008, Gromke and Ruck, 2009, Gromke, 2011)

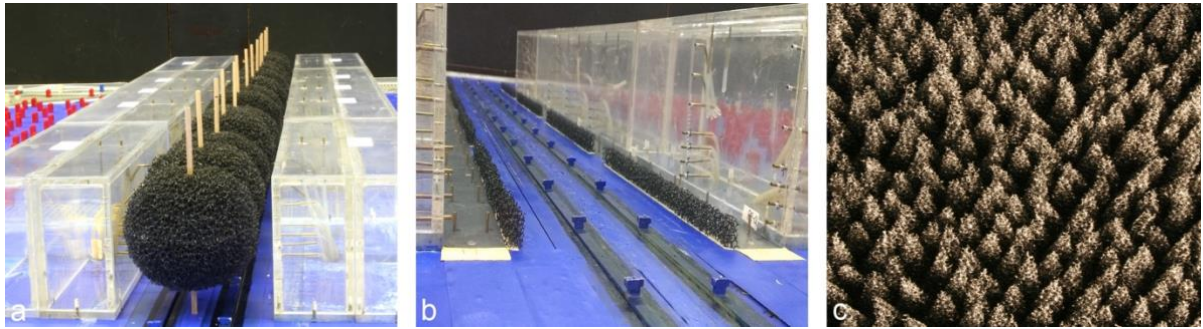

**Fig. 3**. Application examples of open porous foam processed to model (a) avenue-trees in a urban street canyon (Gromke and Ruck, 2007), (b) hedge-rows in a urban street canyon (Gromke et al., 2016), (c) conifer trees of a forest stand (Gromke, 2018, Gromke and Ruck, 2018).

In order to characterise the permeability for airflow of a porous medium in an aerodynamic manner, the pressure loss coefficient can be employed. The pressure loss coefficient $\lambda$ of a porous sample in forced-flow is defined as (Gromke, 2011):

$$\lambda = \frac{\Delta p_{st}}{p_{dyn} d} = \frac{p_{ww} - p_{lw}}{0.5 \rho U_f^2 d} \quad (2.2\text{-}1)$$

with $\Delta p_{st}$ the difference in static pressure between windward (subscript ww) and leeward (subscript lw) of the porous sample, $p_{dyn}$ the dynamic pressure, $\rho$ the fluid density, $U_f$ the bulk flow speed, and $d$ the porous sample thickness in streamwise direction. The $\lambda$-values of foams with identical PPI-value from various manufactures and also among production batches may vary to some extent. As an indication $\lambda_{PPI-7} = 200$ m$^{-1}$, $\lambda_{PPI-10} = 250$ m$^{-1}$, $\lambda_{PPI-20} = 500$ m$^{-1}$, and $\lambda_{PPI-30} = 1000$ m$^{-1}$, which cover most of the required permeability of model vegetation in fluid tunnel studies, can be adopted (Gromke et al., 2016, Klausmann and Ruck, 2017).

The similarity in regard to aerodynamic resistance is ensured if the ratio of momentum extraction $F_d$ by the (model) vegetation to the momentum of the undisturbed approach flow $F_f$ is equal in model-scale and full-scale, i.e $[F_d/F_f]_{ms} = [F_d/F_f]_{fs}$, where subscripts ms and fs stand for model-scale and full-scale, respectively. As is shown in Gromke (2011) and Gromke (2018), by employing Eq. (2.2-1), the following relationship can be derived

$$\frac{\lambda_{fs}}{\lambda_{ms}} = \frac{d_{ms}}{d_{fs}} = M \qquad (2.2\text{-}2)$$

with $M$ a geometric scale factor. Eq. (2.2-2) is the scaling relation which links the aerodynamic resistance, expressed by the pressure loss coefficient, of model and real vegetation. It states that the pressure loss coefficient of the model vegetation is that of the real vegetation divided by geometric scale factor. For pressure loss coefficients of real vegetation the reader is referred to the work of Grunert et al. (1984). Herein, pressure loss coefficients for various trees and shrub species are given as $1.8 \text{ m}^{-1} < \lambda_{fs} < 6.9 \text{ m}^{-1}$ at a wind speed of 4 ms$^{-1}$ and as $0.8 \text{ m}^{-1} < \lambda_{fs} < 3.5 \text{ m}^{-1}$ at a wind speed of 11 ms$^{-1}$.

The proposed scaling and similarity concept complies with the requirements towards drag and permeability as outlined in the previous section on fundamental considerations. The modelling approach including the scaling and similarity concept is, next to its application in reduced-scale wind tunnel studies, also applicable in investigations with scaled vegetation in water channels.

### 2.2.3 Recent advances and outlook

Studies with model vegetation in fluid tunnels contributed to our understanding in the areas of flow and turbulence in and above forest canopies, (e.g. Meroney, 1968, Sadeh et al., 1971, Chen et al., 1995, Novak et al., 2000, Marshall et al., 2002, Morse et al., 2002, Ruck et al., 2010, Tischmacher and Ruck, 2013, Conan et al., 2015), wind loads on trees and storm stability of forest stands (e.g. Stacey et al., 1994, Gardiner et al., 1997, Marshall et al., 1999, Marshall et al., 2002, Gardiner et al., 2005, Tischmacher and Ruck, 2013), exchange and deposition of scalar species and pollutants in forest canopies (e.g. Meroney, 1970, Ruck and Adams, 1991, Aubrun and Leitl, 2004, Aubrun et al., 2005, Wuyts et al., 2008, Conan et al., 2015, Coudour et al., 2016), wind energy-related subjects at forest sites (e.g. Sanz Rodrigo et al., 2007, Desmond et al., 2014, Desmond et al., 2017), and on windbreak by vegetation shelterbelts (Guan et al., 2003, Bitog et al., 2011).

The vegetation modelling approach described in this contribution was successfully applied in studies of the effect of avenue-trees and hedge-rows on flow and pollutant dispersion in urban street canyons (Fig.3a, b) (Gromke, 2011, Gromke and Ruck, 2012, Gromke et al., 2016) as well as flow above forest canopies and wind loads on trees in forest stands (Fig.3c) (Gromke and Ruck, 2018). Next to their contribution to fundamental knowledge, the data of these studies widely serve for validation of numerical flow simulations by computational fluid dynamics (CFD). In particular, the modelling concept described herein is due to its parametrization straightforwardly applicable or implementable in CFD (e.g. Balczó and Ruck, 2009, Buccolieri et al., 2009, Salim et al., 2011, Moonen et al., 2013, Gromke and Blocken, 2015, Jeanjean et al., 2015, Vranckx et al., 2015, Morakinyo and Lam, 2016, Merlier et al., 2018, Moayedi and Hassanzadeh, 2022, Zhu et al., 2022).

Future advancement in modelling of vegetation in fluid tunnels may envisage fluid-structure interactions typically occurring at moderate and higher flow speeds (Stacey et al., 1994, Hao et al., 2020). Moreover, most of the model vegetation models utilized in the past and current investigations do not, or only partly, reproduce reconfiguration and streamlining. The associated aerodynamic effects, such as changes in permeability and reduction of drag coefficient with increasing flow speed, are in general not sufficiently represented (Manickathan et al., 2018). Future fluid tunnel studies in the nexus of vegetation and urban climate may address the effects of e.g. façade or roof greening and parks or green spaces on urban flows with their implications for air quality, natural ventilation, and cooling (Li et al., 2022a, Manickathan et al., 2022, Zhao et al., 2023).

## 2.3  Modelling of solar radiation
### 2.3.1  Fundamental considerations
Short- and long-wave radiation are major contributors in the energy balance at urban and building surfaces. As illustrated in Fig. 4a, short-wave radiations are mainly assumed to be a heat source on impacted surfaces. On the other hand, long-wave radiation in an urban context concerns radiative exchanges between a specific surface and other urban surfaces, ground, and sky dome (Energy, 2018). Radiative fluxes in combination with convective fluxes over the external surfaces result in non-isothermal conditions in street canyons. This phenomenon is mainly presented by an applied heat from heat sources in fluid tunnel studies (Cui et al., 2016, Gong et al., 2022).

To more realistically model both radiation and convective fluxes and avoid difficulties in the implementation of heated surfaces by artificial heaters, one can argue that using a solar simulator can technically present the radiative fluxes in fluid tunnels. Solar simulators have been used in many industrial applications, ranging from vehicle producers to photovoltaic manufacturers (Gallo et al., 2017, Li et al., 2022b). Nonetheless, studies to simulate radiation in atmospheric fluid tunnels are rare in literature. This is again due to the difficulties to provide a consistent and uniform radiation intensity on the impacted surfaces in addition to the blockage that a solar simulator might create against the working fluid (Mirzaei et al., 2014).

### 2.3.2  Recommendations for the design of fluid tunnel experiments
Simulations of solar radiation within fluid tunnels face a wide range of challenges and barriers, which hinders a widespread adoption of such studies (Mirzaei and Carmeliet, 2015). Nevertheless, the conducted studies pave the way for future research to benefit from a combined observation of radiative and convective fluxes in fluid tunnels.

When experiments are designed to integrate solar simulators into the fluid tunnels, a range of considerations should be taken into account. In terms of safety and hazard prevention of the fluid tunnel experiment, materials placed against solar simulators to absorb the radiation intensity should be cautiously selected to avoid melting and fire problems when a constant radiative flux can cause a sudden increase in the surface temperature of the exposed surfaces. While the excess surface temperature may not occur and therefore be noticed in the higher airflow velocities, this can be the case when the operating velocity in a fluid tunnel is considerably decreased. Hence, ensuring a safe range of operating temperature for the exposed surfaces can be initially tested when the wind tunnel is turned off and the operating velocity is zero. The wiring and installation of lamps also should follow the existing safety protocols to prevent any electrocution risks and fire hazards.

In terms of technical challenges, the choice of hot-wire anemometer, the place of the probe and the way it is protected against the radiation are of paramount importance. Probe surfaces can be covered with aluminium foils, and this can be an effective strategy for other building model surfaces manufactured with materials of low melting points. Moreover, solar simulators generate the radiative flux with one or multiple lamps, which should be selected based on the needs of an experiment (Tawfik et al., 2018).

Placing one or multiple lamps as a solar simulator may cause a nonuniform heat flux at the target surface. This is due to the shape of the lamp units even though a more effective design can reduce this discrepancy. In general, the radiation intensity can be expected to be uniform at the middle of the target surface, but less uniform on the edges. Thus, it is essential to monitor the uniformity of radiation at different points of a target surface with the related sensors such as Thermopiles (Renné, 2016) before starting experiments to ensure that the variation range is not more than few percent.

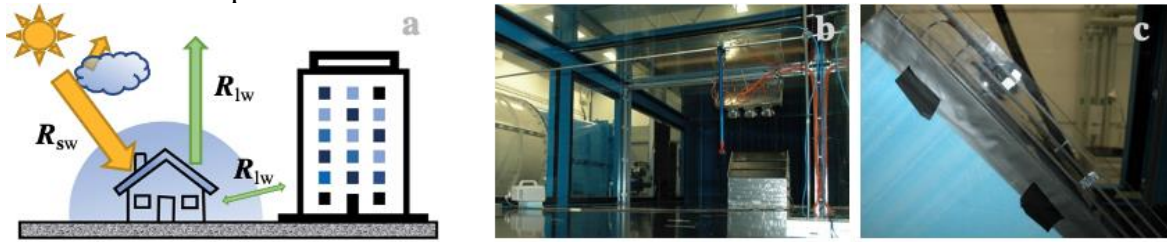

**Fig. 4**. (a) Short-wave and long-wave radiation over building surfaces; (b) and (c) employment of a solar simulator in a wind tunnel (Mirzaei and Carmeliet, 2015)

*2.3.3 Recent advances and outlook*

A solar simulator installed in an atmospheric fluid tunnel can be combined with the advanced techniques such PIV to observe the flow above and within a cavity behind a building integrated photovoltaic (BIPV) (e.g., see Fig. 4b, c). In such studies, the laser beam should penetrate through an unblocked pathway of objects to enable visualizing the flow in the cavities. It should be noted that the PIV technique has a relatively small laser beam area, which might not be large enough to observe the whole domain of the flow. In such cases, focusing the PIV in multiple smaller planes and combining them can be a potential solution, especially when the use of hot-wires in small areas could disturb the flow (Mirzaei et al., 2014). While working with a laser beam demands its own health and safety considerations and training, designing a pathway for the laser sheet to reach cavities and enclosed spaces is an essential part of the experiment with suitable transparent materials such as glass or Plexiglas. Note that a high-speed camera also needs to be installed in a way to have a clear focus over the monitored cavity. The experiment should also ensure that the flow reaching the cavities contains enough seeds to be illuminated by the laser beam to visualise the flow field.

As another advanced technology, thermography techniques are used in different studies to observe the surface temperatures in large and atmospheric fluid tunnels (Le Sant et al., 2002, Mirzaei and Carmeliet, 2015). In atmospheric fluid tunnels it is crucial to ensure that the calibration of infrared cameras follows the standard operating procedures (Martin et al., 2022). In such experiments, infrared cameras should be preferably mounted in the upstream direction within a fluid tunnel to avoid disturbance to the flow while not being optically impacted by the fluid tunnel's reflecting surfaces, e.g., plexiglass.

**2.4 Modelling of thermal stratification**

*2.4.1 Fundamental considerations*

Stable stratification is a common phenomenon in the ABL, which affects the land-atmosphere interactions in terms of heat, mass and momentum transfer processes (Lee, 1979). The stability of the ABL is quantified with the buoyancy frequency (also called Brunt-Väisälä frequency) (Stull, 1988), *N*, which is defined in Eq. (2.4-1) as follows.

$$N = (g\beta\, \partial\theta/\partial z)^{1/2} \qquad 2.4\text{-}1$$

where $g$ is the gravitational acceleration, $\beta$ is the fluid thermal expansion rate, $\theta$ is the potential temperature in the vertical boundary layer, $z$ is the vertical coordinate, and $\partial\theta/\partial z$ is the potential temperature gradient. In the ABL, the typical value of $N$ is around 0.015 s$^{-1}$ (Hunt et al., 1988, Reuten, 2006). Three physical processes can cause stable stratification (Largeron and Staquet, 2016, Czarnecka and Nidzgorska-Lencewicz, 2017, Ning et al., 2018, Niedźwiedź et al., 2021). The first one is a warm front over a cold front, which causes a strong temperature gradient at the interface of the two fronts and is also named elevated inversion. The second one is due to hot air subsidence at a certain location caused by the large-scale (regional scale or global scale) atmospheric circulation. The third one is radiative cooling of land surfaces at night, which causes surface-based inversion and is the most common type in diurnal cycles. The inversions in the ABL can be caused by the joint effect of the above three physical processes. The accumulation of cold air in basins or valleys can form cold pools (Clements et al., 2003, Princevac and Fernando, 2008, Vosper and Brown, 2008, Lareau et al., 2013, Yu et al., 2017) and creates extremely strong inversion ($N$ is as high as 0.1 s$^{-1}$).

To simulate the flow phenomenon in a stable ABL, a temperature gradient or density gradient also needs to be created in fluid tunnels. The non-dimensional parameters to guarantee similarities between the prototypes and reduced scale models are $Fr$ (Lu et al., 1997b, Cenedese and Monti, 2003, Fan et al., 2018, Fan et al., 2020, Yin et al., 2020) or $Ri$ (Ogawa et al., 1985, Uehara et al., 2000, Zhao et al., 2022a).

### 2.4.2 *Recommendations for the design of fluid tunnel experiments*

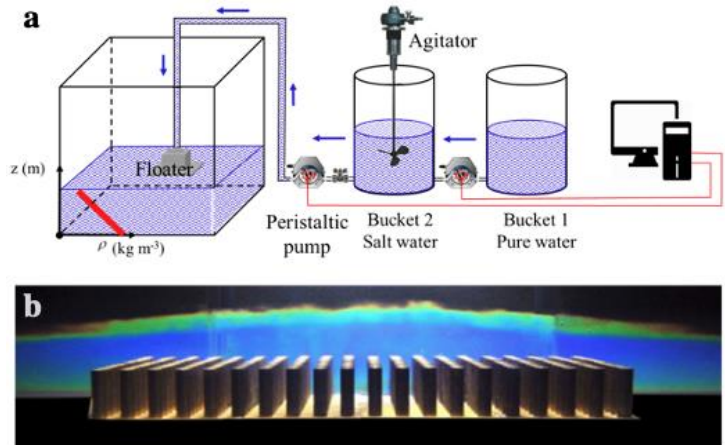

**Fig. 5**. (a) Illustration of the procedure of creating the stable stratification with salt water. Modified from Fan et al. (2018). (b) Temperature field, presenting a dome shape, over a reduced-scale city model in a water tank under stable stratification condition, which is visualized by thermochromic liquid crystal sheet (Fan et al., 2017).

There are two main methods for generating stable gradients, i.e., one is using a temperature gradient and the other one is using a density gradient (such as salt water). In wind tunnels, the temperature gradient can be built up by cooling the bottom and heating the top (Ogawa et al., 1981, Guo et al., 2021), which is also applicable in water tank experiments (Lu et al., 1997a, Lu et al., 1997b). Stable stratification created with salt water can be achieved in water tank experiments with the two-bucket filling technique (Fan et al., 2018, Fan et al., 2019), as shown in Fig.5a. There are several advantages of using salt water. First, the density gradient can be much larger than that with a temperature gradient method. Second, the density gradient will last a longer time than the temperature gradient and does not require a thermal insulation of water tank walls and bottoms. Third, the heating at the top can block the laser or light access when applying PIV to measure the velocity field, which is not a problem in setups with salt water stratifications. Fourth, density gradient profiles can be flexible by adjusting the filling speed of salt water and pure water, for example, a neutral layer covered by a stable layer or a stable layer covered by a neutral layer (Fan et al., 2021a). If the heating method is adopted, only a linear gradient can be achieved as it utilizes heat conduction to build up the temperature gradient.

*2.4.3 Recent advances and outlook*

As the frequency of heatwaves and calm weather conditions increases, the buoyancy-driven flow, such as an urban heat dome in Fig. 5b, becomes more and more important due to their impact on pollutants and heat dispersion at building and city scales (Zhao et al., 2020, Fan et al., 2021b). When the buoyancy-driven flow is dominant or at least comparable to the approaching flow, the similarity is easier to be achieved in water channels than that in wind tunnels. Moreover, as city size increases, the Coriolis force, which is quantified by the Rossby number ($Ro$) (Warn et al., 1995, Embid and Majda, 2006, van der Laan et al., 2020), starts to play a role in modulating the buoyancy-driven flow over urban areas. In this case, a rotating water tank would be a useful tool for simulating Coriolis force related large-scale flow structures at the city scale under Coriolis force.

## 2.5 Modelling of pollutant dispersion

*2.5.1 Fundamental considerations*

Pollutant dispersion in urban areas is driven by local meteorological conditions and the urban morphology. Furthermore, the multiplicity of pollutants and sources broadens the range of parameters that affect the phenomenon. The ability to control the different variables individually and to isolate their effects is the main advantage of dispersion studies in fluid tunnel experiments.

Similarity in laboratory models for pollutant dispersion in urban areas is ensured by means of geometric similarity, and the matching of the main dimensionless numbers for the flow field: $Re$, $Pr$, densimetric $Fr$ and $Sc$ numbers (Snyder, 1981, Meroney, 2004, Tominaga and Stathopoulos, 2016, Mei et al., 2023). Moreover, specific similarity criteria concerning the pollutant emission, should be matched, including the geometric similarity of the source, $Re$ and $Fr$ at emission location, and the density and speed ratios at emission location (Pournazeri et al., 2012, Marro et al., 2014). These parameters are fundamental for simulating the release of non-neutrally buoyant plumes from elevated sources within the city. To reproduce vehicle exhausts, the buoyancy effects and the emission speed are generally neglected, and the similarity is attained using a tracer with density similar to that of the fluid. In this case, however, traffic-induced turbulence has non-negligible effects on pollutant dispersion and the similarity criterion by Plate (1982) can be applied (Gromke and Ruck, 2007).

*2.5.2 Recommendations for the design of fluid tunnel experiments*

Pollutant dispersion in urban areas has been studied in both wind tunnels and in water tunnels. Besides the urban geometry, a key component of the experimental setup is the emitting sources which are generally elevated or ground-level point sources reproduced via metallic tubes, or line sources to simulate exhausts from vehicles. The latter is designed to minimize the vertical momentum and maximize lateral homogeneity (Meroney et al., 1996). Above the line source, traffic-induced turbulence can be mimicked by plates mounted on rotating bells (Kastner-Klein et al., 2001), as shown in Fig. 6.a.

To simulate neutrally buoyant emissions in wind tunnel experiments, a mixture of hydrocarbon with air is generally injected (Yee et al., 2006, Salizzoni et al., 2009, Perry et al., 2016). Point concentration measurements are achieved with a Flame Ionization Detector (FID). Alternatively, sulfur hexafluoride is used as a tracer gas (Yassin et al., 2005, Gromke and Ruck, 2007, Chavez et al., 2011), which can be collected and sent to the detector via a capillary tube or taken from measurement taps at building walls. Water vapor produced by a $H_2O$ atomizer and measured by humidity sensors has also been used as a tracer for dispersion over urban areas (Mo and Liu, 2018).

Buoyant plume emissions (Fig. 6.c) in wind tunnels are reproduced by means of light or heavy gases (He, $CO_2$) or heated air (Robins et al., 2001, Snyder, 2001, Kanda et al., 2006). In the first case, a tiny quantity of a gas tracer detectable by a FID is generally added to the buoyant gas (Vidali et al., 2022). In the second case, measurements are performed with standard thermocouples (Marro et al., 2014). Beside gaseous sources, Rodriguez et al. (2020) measured the concentration of ultrafine particles in the wake of vehicles. In water channels, fluorescent dyes are released as passive tracers and the LIF technique allows for simultaneous multipoint concentration measurements (Yee et al., 2006, Wangsawijaya et al., 2022) (Fig. **6**.6 b). Buoyancy effects can be obtained by mixing tracer dyes with alcohol and water or releasing salt water (Pournazeri et al., 2012).

To evaluate chronic pollution in urban areas, it is often sufficient to estimate the average pollutant concentration over time. To ensure a reliable prediction, the acquisition and averaging time has to be longer than the typical time scale of the vortical structures within the domain (Pavageau and Schatzmann, 1999, Garbero et al., 2010). Conversely, the assessment of hazards due to toxic or explosive pollutants, or the impact of odours, requires the analysis of concentration fluctuations (Cassiani et al., 2020) from which higher-order concentration statistics and probability density functions can be determined (Gailis and Hill, 2006, Yee et al., 2006, Klein et al., 2011). Velocity and concentration must be simultaneously measured to estimate turbulent mass fluxes, which are key for understanding pollutant exchange in complex geometries (Carpentieri et al., 2012, Marro et al., 2020).

*2.5.3 Recent advances and outlook*

In the last decades, physical models in fluid tunnels have brought great advances in understanding the mechanisms of dispersion in urban areas at different scales (Britter and Hanna, 2003, Xia et al., 2014, Zhang et al., 2020). For a group of sparse obstacles in the wake regime (Fig. 6d), the planar and frontal area density of buildings are found to affect the dispersion process, and the concentration profiles within the building array are in good agreement with Gaussian plume models (Davidson et al., 1996, Macdonald et al., 1998). This behaviour differs in realistic and dense urban geometries (e.g. Garbero et al., 2010), where the decoupling between the layer above the roofs and the region within the streets leads to channelling effects, deflection of plume centreline and complex horizontal spreading. As regards the effect of non-neutral approaching ABL winds on a regular array, the average concentrations in the canopy can be up to two times higher in stable stratification than in the neutral case and three times lower in convective conditions (Marucci and Carpentieri, 2020a) Much effort has been devoted to the understanding of dispersion in a single street canyon and at street intersections (Ahmad et al., 2005, Yazid et al., 2014). Recent research has shown how the warming of the walls can lead to the deterioration or improvement of air quality depending on the canyon aspect ratio (Marucci and Carpentieri, 2019, Fellini et al., 2020). Different studies (Hajra and Stathopoulos, 2012, Nosek et al., 2016, Llaguno-Munitxa et al., 2017) have shown how the shape of buildings and roofs (Fig. 6e) have a non-negligible effect on the concentration in the streets. Recently, a growing interest has been devoted to the effect of tree planting (Fig. 6f) (Gromke and Ruck, 2007, Gromke and Ruck, 2009, Gromke and Ruck, 2012) on pollutant dispersion in a single canyon.

To face the challenges related to climate change and urbanization, a greater number of experimental studies on the effect of vegetation and solar heating on air pollution is crucial. Further experiments on the dispersion of ultrafine particles in the wake of vehicles would help considerably to characterise particle exposure at pedestrian level, though care must be given to the similarity criteria for particles. This review also reveals the lack of experimental data on the concentration of reactive plumes in urban geometries that would be fundamental to validate the large number of numerical models covering the topic.

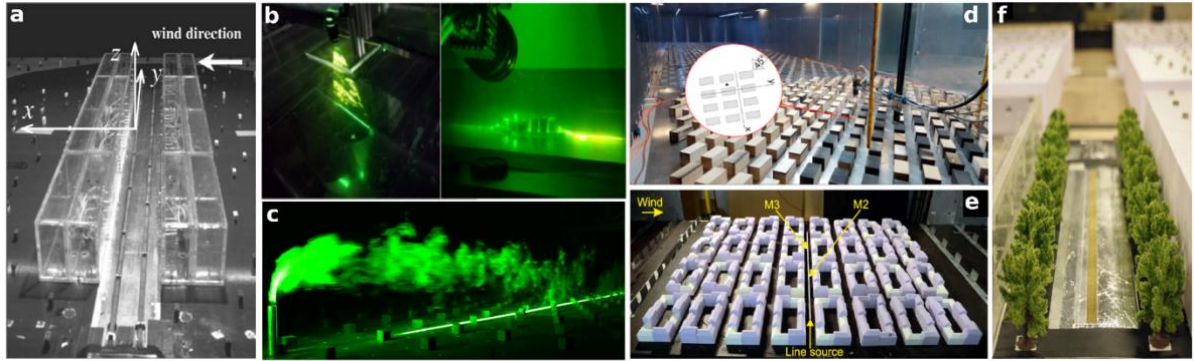

**Fig. 6**. Fluid tunnel testing of pollutant dispersion in cities: (a) Moving belts along a street canyon to simulate two-lane traffic (Kastner-Klein et al., 2001) (b) PIV-PLIF setup in a water flume (Lim et al., 2022). (c) Laser tomographic visualisation of a heavy gas plume (Vidali, 2021). (d) Point source (Marucci and Carpentieri, 2020a) and (e) line (Nosek et al., 2016) source at street level to simulate emissions in an urban array. (f) Street-level linear source to mimic pollutant dispersion in a vegetated canyon (Fellini et al., 2022).

### 2.6 Modelling of indoor and outdoor natural ventilation
*2.6.1 Fundamental considerations*

Natural ventilation occurs due to pressure differences arising naturally between openings in a building, which drive the exchange of air between indoor and outdoor spaces. The pressure differences can be caused by two main mechanisms: wind and buoyancy effects (or a combination of both). In both cases, the flow regime that controls the exchange of air between indoor and outdoor is largely dependent on the location and geometry of the openings, pressure and temperature boundary conditions at the building envelope and the presence of local buoyancy sources (Linden, 1999).

Pressure distributions due to wind are generally assessed through wind tunnel experiments. The nature of the urban environment and the building openings, with all their sharp edges makes wind speed largely irrelevant due to flow separation. This aspect simplifies experimental investigation as it makes the problem essentially *Re*-independent (Linden, 1999).

On the other hand, buoyancy-driven flows due to temperature differences may represent a challenge for fluid tunnel modelling, due to the lower *Re* leading to an increased importance of viscous effects (Linden, 1999). The buoyancy force can be described in terms of the reduced gravity, $g'$:

$$g' = g\frac{\Delta \rho}{\rho} = g\frac{\Delta T}{T} \qquad (2.6\text{-}1)$$

where $g$ is the acceleration of gravity, $\rho$ the density and $T$ the temperature. The dimensionless number of concern are the Reynolds number ($Re$) and the Peclet number ($Pe$), which can be written in terms of the reduced gravity as:

$$Re = \frac{\sqrt{g'H}\,H}{\nu}; \quad Pe = \frac{\sqrt{g'H}\,H}{\kappa} \qquad (2.6\text{-}2)$$

where $\nu$ is the kinematic viscosity, $\kappa$ is the coefficient of molecular diffusivity, and $H$ is the relevant vertical scale. In order to reduce the mismatch in $Re$ and $Pe$, small-scale experiments in water (e.g., using salinity to simulate buoyancy) are generally used (Linden et al., 1990, Davies Wykes et al., 2020). Fluid tunnel experiments can still be of value in buoyancy-driven flow, as specialized facilities might help in characterising temperature and heat exchange around buildings in urban areas, e.g. Marucci and Carpentieri (2019).

*2.6.2   Recommendations for the design of fluid tunnel experiments*

Measurements of ventilation rates in a wind tunnel are generally done using tracer concentration techniques (Etheridge, 2011). This would require an injection of a tracer gas in the building of interest and measuring the concentration decay due to ventilation. A fast response instrument is then needed to capture the transient and typically this is done through a Fast-response Flame Ionisation Detector (FFID), e.g., Marucci and Carpentieri (2020a) used hydrocarbons as tracer gases. FFID can also be used for air quality studies when assessing interactions between the interior and the exterior of the building for pollution dispersion purposes.

Creating the correct wind environment is relatively easy in large boundary-layer fluid tunnels, where neutral conditions can be easily achieved in urban wind flows. Specialized facilities on the other hand, are required if non-neutral (stable or convective) boundary layers are to be generated (Marucci and Carpentieri, 2020b). As mentioned in the previous section, correct scaling of combined wind- and buoyancy-driven flows can be particularly challenging as temperature differences between indoor and outdoor environments might have to be greater than 75ºC (Etheridge, 2011).

Conventional techniques (e.g. laser doppler anemometry, LDA, or PIV, can be used to characterise boundary conditions around buildings. Direct pressure measurements are less frequent as characterising pressure distributions with a high resolution on building surfaces can be challenging, especially when pressure values are small (Nathan et al., 2021).

*2.6.3   Recent advances and outlook*

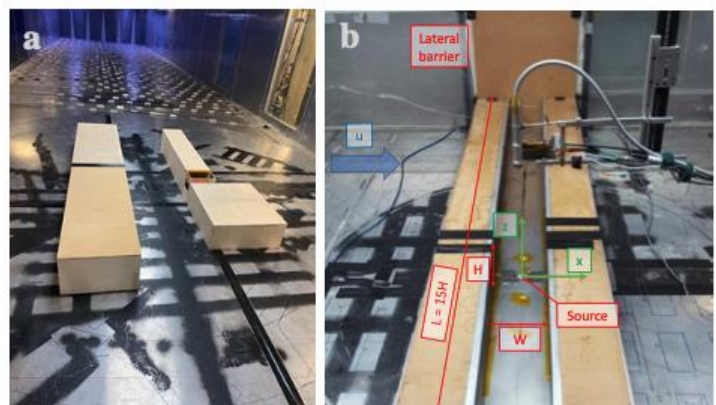

**Fig. 7.** (a) Study of combined wind- and buoyancy-driven ventilation and (b) study of the effects of local buoyancy sources on flow and dispersion in street canyons, from (Marucci and Carpentieri, 2019); both studies from the EnFlo wind tunnel, University of Surrey, UK.

Simulating natural ventilation in a fluid tunnel has great potential, but also several limitations, as explained above. Indoor and outdoor studies have traditionally been carried out independently, usually for different purposes, and only recently some research projects have started the connection between the two environments, see, for example, the MAGIC project by Song et al. (2018) in Figure 7a.

Thermal characterisation of the building boundary conditions cannot be achieved easily in non-specialised facilities, but some specialised ones are starting to bridge the gap, with studies on non-neutral urban boundary layer (UBL) flows (Marucci and Carpentieri, 2020a, Marucci and Carpentieri, 2020b) and local heating sources (Marucci and Carpentieri 2019; see Fig. 7b). Recent advances include the development of post-processing techniques for measuring pressure and pressure fluctuations in low-speed wind tunnels, where the pressure variations are very small (Nathan et al., 2021).

## 2.7 Modelling of outdoor wind thermal comfort
### 2.7.1 Fundamental considerations
Thermal comfort is a subjective evaluation of the thermal environment. Over the last century, researchers have created several thermal comfort models to predict human thermal responses to different combinations of environmental parameters (including air temperature, humidity, radiation, and wind velocity) and personal variables (e.g., clothing and activity levels) (Gagge et al., 1972, Tanabe et al., 2002, Fiala et al., 2012, Chew and Norford, 2018). These models mathematically describe the thermoregulation processes, in which convective heat transfer coefficient (CHTC) is required to estimate the heat exchange between the human body and its surrounding environment. To measure the anatomically specific CHTCs for individual body segments, thermal manikin experiments are conducted in fluid tunnels, which features the ability to generate a wide range of wind conditions in a limited time and to maintain the same wind condition for a manikin to reach a steady state. The dimensionless numbers to describe force convection are *Nu*, *Re*, and *Pr* (De Dear et al., 1997). Although a full-scale thermal manikin is used in most of the studies, the reduced-scale models may be the focus of future modelling, since they can significantly reduce the sampling time under each test condition.

### 2.7.2 Recommendations for the design of fluid tunnel experiments
A thermal manikin has been proven as an effective instrument for measuring sensible heat transfer between the human body surface and the surrounding environment (Tanabe et al., 1994, De Dear et al., 1997). It features an anatomically realistic human morphology, with a precision heating element and temperature sensor system embedded within the "skin".

For the approaching flow, in addition to mean wind velocity, turbulence intensity and power spectral density distribution also play a key role in determining thermal perception and, therefore, the pedestrian level turbulence characteristics need to be properly simulated for thermal comfort studies in fluid tunnel modelling (Zhou et al., 2006).Turbulence generators, such as oscillating aerofoils and passive grids (Fig. 8b) are implemented upstream to simulate the full-scale natural wind. The approaching wind profile or the wind profile close to the body segments needs to be measured using fast-response anemometers.

The manikin is often placed at the centre of the fluid tunnel test section, in a sitting, standing, or walking posture (Fig. 8 a-c), and tested under target wind conditions. CHTC is calculated based on the thermal state (skin temperature and heat loss) automatically logged by the manikin itself, and the air temperature and wind tunnel inner surface temperature can be measured by additional temperature sensors (Fig. 8a). The relationship between CHTC and the approaching wind condition is generally established through empirical regressions.

To further investigate pedestrians' physiological and perceptual responses to wind, human subjects are invited to the wind tunnel to experience different wind conditions. Their thermal physiological parameters (e.g., local skin temperature) and perceptual responses can be collected and compared with thermal comfort model output (Yu et al., 2021).

*2.7.3   Recent advances and outlook*

Thermal manikin and human subject experiments in fluid tunnels enable us to quantify the impact of wind on thermal perception. The CHTCs for individual body segments and whole-body are generated for typical outdoor wind conditions, including wind speed from still air to around 13 m s$^{-1}$ (Li and Ito, 2014), turbulence intensity from 0 to around 30% (Yu et al., 2020), evenly spaced horizontal directions, and different body postures, including sitting, standing, and walking. The prevailing thermal comfort models should adjust the CHTC formula according to the target environments and activity conditions to improve the prediction accuracy.

While the effects of wind on sensible heat loss have been thoroughly investigated in the literature, few studies have focused on evaporative heat loss, which plays a role in determining thermal comfort when subjects' sweat accumulation increases (Bakkevig and Nielsen, 1995). To better understand the impact of body motion on CHTCs, the flow field around the manikin or human subjects during activities such as walking, and cycling is worthy of further investigation (Luo et al., 2014). Also, the impact of relative humidity on thermal perception should be studied in a conditioned wind tunnel, where relative humidity can be controlled to simulate transpiration and sweating.

Based on the field measurement in urban areas, the pedestrian level turbulence intensity ranging between 10 to 60% is not uncommon (Murakami and Deguchi, 1981, Tse et al., 2017, Zou et al., 2021), and about half of the energy of the turbulence concentrates at frequencies less than 0.1 Hz (Hunt et al., 1976). More work needs to be done on simulating in the fluid tunnel a full-scale wind profile with high turbulence intensity and low frequency random gustiness. Future research efforts could also be applied to the interaction effect between wind and radiation by introducing solar simulators into the fluid tunnel.

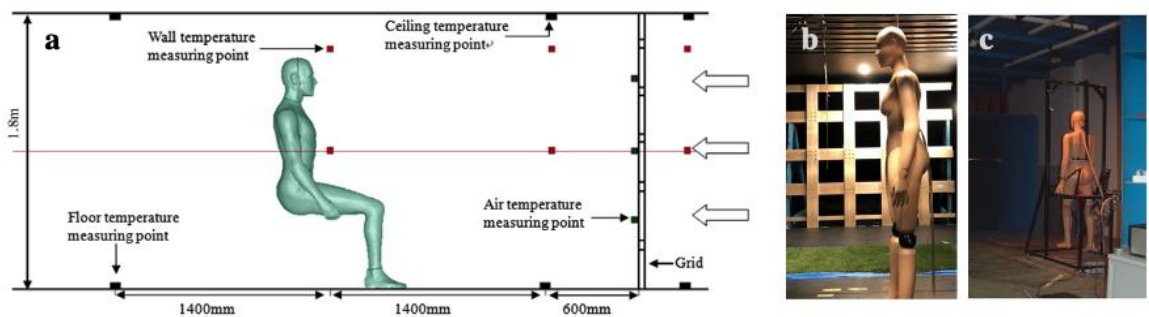

**Fig. 8**. Wind tunnel testing on convective heat transfer coefficient: (a) a schematic diagram of wind tunnel setup and a thermal manikin in sitting posture, modified from Li and Ito (2014); (b) a thermal manikin in standing posture downwind of a passive turbulence generator (Yu et al., 2020); (c) a thermal manikin in a walking posture, modified from Oliveira et al. (2014).

## 2.8  Modelling of urban flow over complex urban sites
*2.8.1   Fundamental considerations*

The use of the fluid tunnels to test urban flow at complex urban sites or a small portion of these has a long heritage. Probably the first experiments inspecting the effect of adjacent buildings were carried out by C. L. Harris in 1934 on the Empire State Building (New York, USA) at the wind tunnel facility of the Bureau of Standards (Harris, 1933). The tests of Harris followed up a series of tests previously performed by H.L. Dryden and G.C. Hill, who investigated the wind pressure on the Empire State Building model not including the surrounding buildings (Dryden and Hill, 1933). Since then, an increasing number of studies have been carried out not only on isolated buildings/structures but also on realistic urban models, leading to a significant improvement in wind tunnel facilities, measurement techniques and the overall scientific background (Fig.9).

In 2022, despite the different eras and the progress made on numerical (e.g. CFD simulations) and field measurements (e.g. anemometers and LiDAR profilers) over different decades, the use of fluid tunnel for testing realistic, complex urban sites is still remarkably high. This is justified by the fact that academicians as well as practitioners typically find with such tests practical and reliable solutions for a large variety of topics of urban physics and wind engineering field, such as the high pollutant dispersion, pedestrian-level wind discomfort, indoor/outdoor thermal comfort, wind energy, and wind loading (Davenport, 2002, Baker, 2007, Moonen et al., 2012, Blocken, 2014, Stathopoulos et al., 2018, Weerasuriya et al., 2018, Simiu and Yeo, 2019, Solari, 2019, Kareem, 2020, Gao et al., 2022).

As mentioned earlier in this paper, a proper downscaling of atmospheric winds and realistic urban models are generally quite challenging. Publications released in the last 50 years have not only further enhanced the scientific background previously developed, but also provided useful practical guidelines to accurately reproduce scaled neutral, stable and unstable atmospheric boundary layer (ABL) winds (Stull, 1988) in fluid tunnel tests (ASCE 49-12, 2012). Important aspects related to similarity criteria (geometric, kinematic and dynamic) to be satisfied between full and reduced scale, both for the ABL flow and urban model, have been extensively investigated. On these grounds, large variety setups of roughness fetch and vortex generators (i.e. spires) to accurately reproduce the turbulent structures of the approach ABL flow upstream of the model have been systematically calibrated and tested in aeronautical, climatic and ABL fluid tunnels (e.g. Jensen, 1958, Castro and Robins, 1977, Cermak, 1981, Irwin, 1981, Saathoff and Melbourne, 1987, Davenport, 1992, Farell and Iyengar, 1999, Plate, 1999, Cermak, 2003, Holmes, 2004). In particular, a set of key parameters has been extensively monitored and found to be crucial for the proper development of a neutral ABL wind along the wind tunnel test-section: mean velocity profiles, stream-wise turbulence intensity, lateral turbulence intensity, vertical turbulence intensity, power spectral density distributions and integral length scales of turbulence (Cermak, 1975). However, the agreement between reduced-scale experimental data and meteorological data from codes and standards (VDI-3783, 2000) still remains an important topic to be addressed to guarantee the validity of the scaled wind characteristics.

*2.8.2 Recommendations for the design of fluid tunnel experiments*
Matching the most relevant dimensionless parameters between reduced scale and full scale can almost never be realised. However, the exact matching of some of these parameters is not that relevant for urban flow studies. As mentioned also in Section 2.6.2, in most cases, buildings of realistic urban models are mainly characterised by sharp edges with fixed separation points of the flow, which makes the problem often *Re*-independent. In addition, although in urban studies the *Re* threshold (for turbulent flows) is typically exceeded, specific tests to investigate the *Re* independence are always highly recommended. The *Ro* and *Fr* which account respectively for the effect induced by the rotation of the earth on the wind (e.g. Ekman spiral) and the gravity effect on the flow pattern are typically irrelevant in urban studies, considering also that the former is almost unrealisable in ordinary fluid tunnels. In contrast, the *Ri*, *Gr* and *Sc* may play a key role in properly scaling thermal effects and dispersion phenomena (see also sections 2.1, 2.4, 2.5, 2.7).

On top of that, there are some other important aspects that may be crucial for the choice of the scale: (i) the fluid tunnel test-section length, necessary to properly develop the approach ABL wind (Cermak, 1975); (ii) the blockage ratio, such as the ratio between the frontal area of the model and the wind tunnel cross-section that should not exceed the 5% (see ASCE 49-12, 2012); (iii) the need to manufacture small-scale architectural features to avoid oversimplifications that might threaten the reliability of the experimental results (Carpentieri and Robins, 2015, Ricci et al., 2017b, Pađen et al., 2022); (iv) the encompassment of a sufficient portion of the environment surrounding the area of interest (Ricci et al., 2022). Hence, it is clear that defying the most appropriate scale of a realistic urban model is ultimately a compromise of multiple factors, which gives rise to a wide range of scales commonly adopted, from 1:200 (for small districts) to 1:1000 (for large portions of cities) approximately. With such small scales, the choice of the materials and the manufacturing technique can also help to improve the resolution of geometries by significantly reducing the gap between reality and models. Geometrical simplifications adopted for buildings or a portion of these might have an influence on the UBL and urban canopy layer (UCL) development but also on the local wind flow pattern (e.g. inside canyon streets) (Ricci et al., 2017a, Ricci et al., 2017b).

The choice of materials is also strictly related to the type of tests to be performed (e.g. wind speed, wind pressure, temperature, pollutant dispersion measurements) and consequently to the instrumentation that should be used (e.g. pressure taps, Irwin probes, cobra probe, hot-wire anemometry, laser-doppler anemometry, particle image velocimetry). In this perspective, due to the small dimensions of buildings and streets, the use of a non-intrusive and accurate measurement technique is highly recommended for urban models. Advantages and disadvantages of measurement techniques can be different when related to specific topics, however, giving an exhaustive overview is beyond the scope of this paragraph. Finally, since these models often extend beyond the fluid tunnel turntable, it is equally important to accurately define the monitored area of the models (where measurements will be performed) preferably far away from the lateral boundaries of the facility by avoiding any possible undesirable interference effects (Stathopoulos and Surry, 1983).

### 2.8.3 Recent advances and outlook

Most advanced techniques like 3D printing and injection moulding can facilitate the manufacture of such complex urban models, improve the geometry resolution, pre-arrange during the design stage of the model for the sensor installation (e.g., pressure taps) to definitively reduce the discrepancies between reality and models. In that regard, the use of numerical techniques (e.g. CFD), as a complementary tool to support the fluid tunnel tests of ABL winds on realistic urban models, might help to gain a better understanding of (i) the size of the surrounding environment to be realised for fluid tunnel tests, (ii) the level of geometrical simplifications to be adopted for buildings and other urban features, (iii) the UBL and UCL development through/over the investigated area.

Nevertheless, the climate change and the increasing number of extreme events, different from most ordinary ABL winds, are boosting efforts in detecting, testing, simulating and modelling thunderstorm outflows and tornadoes worldwide (Hangan et al., 2017, Solari et al., 2020). The use of new fluid tunnel typologies and/or dedicated tornado/downburst simulators able to host also extensive urban models has become reality. However, due to downscaling and similarity issues, most of the studies still focus on empty chambers or isolated structures/buildings. This trend is bound to change and most likely more experimental tests on realistic urban models will be carried out in the near future, in order to make buildings and cities more resilient and less costly against storm wind damages.

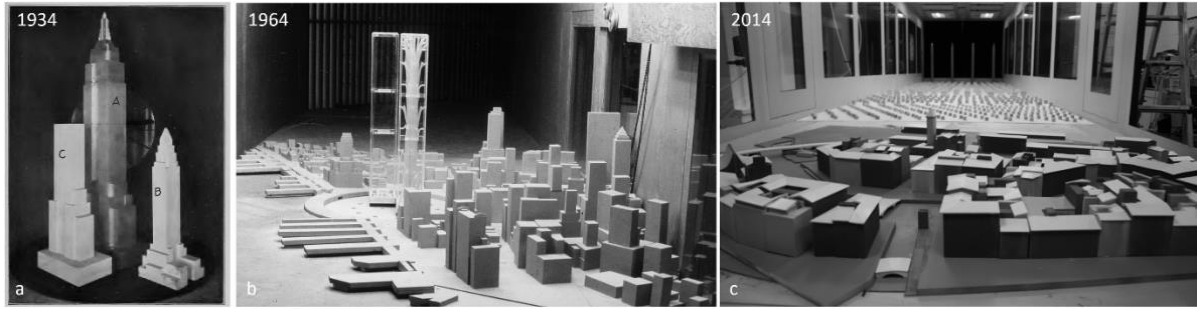

**Fig. 9**. Wind-tunnel testing of parts of realistic urban cities: (*a*) Empire State Building and its immediate surroundings, USA, modified from (Harris, 1933); (*b*) Twin Towers of the New York Trade Center, USA, modified from (Plate, 1999); (*c*) district of Quartiere La Venezia, Livorno city, Italy (with permission of Dr. A. Ricci).

In conclusion, if on one hand the main findings of realistic urban studies are difficult to generalize and be used for basic analytical formulations (as for idealized case studies), on the other hand it has to be acknowledged that probably this is one of the best "trade-off" between academicians and practitioners to gain a reliable understanding of the wind flow field around buildings amidst well-settled urban layouts and provide feasible solutions to actual problems.

## 3. Three challenges for fluid tunnel modelling of the urban climate
### 3.1 Modelling of multi-physics

Multi-physics processes take place in our living cities, such as solar radiation absorption and heat emission from urban materials, convective heat transport by wind, evapotranspiration of plants, evaporation of water bodies, release of anthropogenic heat, emission and dispersion of pollutants, etc. Physical modelling of these processes in fluid tunnels is extremely challenging, if not impossible. The challenges lie in (i) deriving scaling among the multi-physics processes, (ii) complex setups for generating these phenomena simultaneously, and (iii) limited capabilities for large-scale flow, temperature and humidity measurements.

Scaling has been well established for isothermal airflow studies using fluid tunnels where Reynolds number-independency is often assumed (e.g. Chew et al., 2018a, Shu et al., 2020). For buoyancy-involved or non-isothermal airflow, the Richardson number has been accepted as the proper characteristic dimensionless parameter (Aliabadi et al., 2019, Zhao et al., 2022b). However, as to other physical processes, there are few well-established scaling criteria. As an example, scaling for urban surface heat budget that is dominated by shortwave and longwave radiation, convective heat transfer and heat storage by urban materials, has not been established for scaled-down experimental studies. Present studies of these physical processes usually prescribe constant surface temperature or heating capacity to mimic urban heat to some extent. Shading and transpirative cooling by vegetation and plants in urban areas, as another example, could be studied using fluid tunnel experiments based on thoughtful scaling analysis (Manickathan et al., 2022).

In addition, complex experimental setups are needed to reproduce multi-physics processes in a fluid tunnel. Experimental setups in the field of urban climate are often designed to study a particular physical process, rather than to study coupled multiple processes. For studies of urban isothermal wind, fluid tunnel measurements based on scaled-down realistic neighbourhood models have been the norm. This could be the basis where other physical processes, such as heterogeneous urban surface heat budget, could be realised by introducing an artificial solar radiation and selecting building models' materials thoughtfully to mimic heat absorption and emission. In a further step, cooling effects of living vegetation in complex urban streets may be physically modelled using small-sized pot plants.

Last but not the least, substantial development of measurement techniques and advanced post-processing abilities is still much needed for fluid tunnel studies of urban climate. For isothermal airflow, planar- and stereo-PIV have been developed to an advanced level that accurate velocity fields can be obtained. The recent development of water tunnel (flume) PIV-LIF measurement technique provides a way to obtain velocity and temperature fields simultaneously and thus to better understand turbulent and convective heat transport processes (Zhao et al., 2022b). However, for wind tunnel modelling, temperature field measurements still mainly rely on individual thermocouple measurements at certain locations. Instruments that allow air temperature and humidity measurements for a large FOV need to be developed for studies modelling coupled heat and moisture transport in complex urban sites. The development of these instruments should take mobility and flexibility into account to allow efficient measurements with multiple FOVs.

### 3.2 Modelling of anthropogenic processes

Urban climate involves complex interplay of many anthropogenic processes and synoptic climate (Kubilay et al., 2020, Masson et al., 2020). Those typical processes include, but not limited to, emission of anthropogenic heat through air-conditioners, transportation, industrial plants, etc (Mei and Yuan, 2021). The emission of pollutants in the form of aerosols may particularly affect the radiation in the lower atmosphere and even the formation of clouds and precipitation (Masson, 2006, Nazarian et al., 2018). To reproduce urban climate phenomena in fluid tunnels, stochastic and anthropogenic processes in different forms need to be taken into account.

The challenges to reproducing anthropogenic processes in urban climate in fluid tunnels primarily lie in the mimicking of spatial- and temporal-inhomogeneous heat and pollutant sources. As an example, on one hand, to simulate the impacts of spatially inhomogeneous release of anthropogenic heat in different residential areas, accurately designed heating elements in a fluid tunnel are required, which should be capable of maintaining the desired spatially-varied surface temperatures. On the other hand, those controllable heating elements should be able to reproduce and mimic temporal variation of the release of anthropogenic heat, such as varying operational capacity of Heating, Ventilation, and Air Conditioning (HVAC) systems due to different cooling demands in a day. As another example, to study inhomogeneous air quality in cities or in a neighbourhood, spatial- and temporal- dependent anthropogenic pollutant emission processes should be modelled in fluid tunnels.

How to couple these anthropogenic, stochastic processes to prevailing wind, heat, and moisture transport processes at realistic spatial and temporal scales remains another key challenge for fluid tunnel modelling. It has been well established to model meteorological conditions (e.g. wind) from the statistical point of view of prevailing characteristics. However, when it comes to highly time-dependent or fast evolving anthropogenic processes, multiple time scales have to be considered and realised in fluid tunnel modelling to both characterise prevailing (background) urban climate and timewise (superimposed) characteristics.

Given the complexity in matching various spatiotemporal scales, not to mention the development of case-specific experimental setups, the best use of fluid tunnel modelling to understand stochastic anthropogenic processes may lie in providing benchmark experimental modelling data for validation and calibration of CFD models, which ultimately facilitates numerical studies of the full-scale urban climate. High-quality experimental data from fluid tunnels on anthropogenic heat or pollutant emission processes involving varying intensities of wind flow turbulence and buoyancy is much needed for CFD model development, in particular for turbulence modelling and correct treatment of the buoyancy term.

### 3.3 Combined fluid tunnel, scaled outdoor and field measurements

While full-scale field experiments ($H \sim 10 - 100$ m) provide a reliable way to understand urban climate (Eliasson et al., 2006, Offerle et al., 2007), diurnal cycles of urban thermal environment, including solar radiation, thermal storage by urban materials, vegetation transpiration and others, are hard to model in fluid tunnels. Furthermore, urban geometric layouts and building surface materials in real cities are highly heterogeneous, and thermal boundaries are complicated and usually difficult to quantify.

Scaled outdoor measurements ($H \sim 1$ m) have been verified as a good option to obtain high-quality parametric experimental data under realistic meteorological conditions (e.g., Yee and Biltoft, 2004, Kawai and Kanda, 2010, Chen et al., 2020). One of the paramount advantages of scaled outdoor measurements is the representation of physical processes which may rely on nature, realistic conditions, such as solar radiation. Also, scaled outdoor measurements usually allow models one order of magnitude larger compared to fluid tunnel models, which facilitates the matching of characteristic dimensionless numbers. As an example, compared to fluid tunnel studies in which scaled-down models are in the range of $H \sim 0.1$ m, scaled outdoor models are in the range of $H \sim 1$ m. Building models of complex geometries or living vegetation of different species (e.g., trees and shrubs) may be realised in-situ at measurement site, without the need of additional setup.

As it is difficult or even impossible to satisfy all similarity requirements for scaled experimental studies either in laboratory or outdoor setting, a promising and viable approach for investigating urban airflow and thermal environment may be to rely on the combination of numerical simulations and experiments at various scales, i.e., full-scale field measurements, scaled outdoor measurements and fluid tunnel experiments. Conducting measurements covering this wide range of scales also allows us to gain understanding on the effects of scaling and identify physical processes that are particularly sensitive to scaling.

## 4. Concluding Remarks

In this paper, the capabilities of fluid (wind and water) tunnel for modelling of urban climate on eight important physical processes (i.e., transport of heat in airflow, evapotranspiration and aerodynamic effects of vegetation, solar radiation, thermal stratification of airflow, pollutant dispersion, indoor and outdoor ventilation, outdoor wind thermal comfort, and wind dynamics in complex urban settings) have been reviewed. Fundamental considerations, recommendations for design of experimental modelling, recent advances and outlook have been provided for each topic, which serve as a repository of the state-of-the-art of physical modelling using fluid tunnels.

While substantial advances have been made in modelling of decoupled, individual physical processes, grand challenges ahead lie in (i) physical modelling of coupled physical processes, (ii) mimicking of stochastic anthropogenic heat and pollution processes, and (iii) scaling of the different multi-physics processes. Fluid tunnel modelling of multi-physics processes dominating urban climate, such as the joint effects of evapotranspiration of plants and wind flow in complex and realistic urban morphology, is much needed for understanding the physics and also for validation of advanced numerical models that solve multi-physics problems of urban climate. In addition to the challenges in realising those physical processes in fluid tunnels, establishing proper scaling for scaled-down multi-physics processes and full-scale scenarios is even more challenging, and deserves future research efforts, particularly theoretical analyses.

To tackle these grant challenges, a research consortium that comprises experienced researchers working on different urban climate processes is imperative. The rich expertise of such a research consortium would facilitate the design of advanced experimental setups for generating and studying a combination or full set of those important physical processes. The realisation and outcome of fluid tunnel modelling of multi-physics urban processes would in parallel contribute to validation and enhancement of advanced numerical models for urban climate studies.

## Acknowledgements

We acknowledge the funding support from the Swiss National Science Foundation (Grant 200021 169323), the Fundamental Research Funds for the Central Universities (K20220163).

## CRediT authorship contribution statement

**Y. Zhao**: Conceptualization
**Y. Zhao, LW Chew, Y. Fan, C. Gromke, J. Hang, Y. Yu, A. Ricci, Y. Zhang, Y. Xue, S. Fellini, PA. Mirzaei, N. Gao, M. Carpentieri**: writing - original draft, review & editing
**P. Salizzoni, J. Niu, J. Carmeliet**: writing - review & editing

## Appendix A.

Table A1. The eight research areas covered in this paper and their requirements in fluid tunnel modelling. Y for yes, N for no, O for optional (depending on applications).

| Area of modelling | Section | Requirement | | |
|---|---|---|---|---|
| | | Atmospheric boundary layer | Heated surface or thermal plumes | Scalar transport |
| Thermal buoyancy effects | 2.1 | O | Y | O |
| Vegetation | 2.2 | Y | N | N |
| Solar radiation | 2.3 | O | Y | N |
| Thermal stratification | 2.4 | O | Y | N |
| Pollutant dispersion | 2.5 | Y | O | Y |
| Indoor and outdoor natural ventilation | 2.6 | Y | O | O |
| Outdoor wind thermal comfort | 2.7 | Y | Y | N |
| Urban flow over complex urban sites | 2.8 | Y | Y | Y |